\documentclass{llncs}
\usepackage{graphicx} 
\usepackage{algorithmic}
\usepackage{graphicx}
\usepackage{textcomp}
\usepackage{xcolor}
\usepackage{comment}
\usepackage{url}
\usepackage{amssymb,amsmath}
\usepackage{listings}

\definecolor{dkgreen}{rgb}{0,0.5,0}
\definecolor{gray}{rgb}{0.5,0.5,0.5}
\definecolor{mauve}{rgb}{0.58,0,0.82}
\definecolor{red}{rgb}{1,0,0}
\definecolor{color_type}{rgb}{0.0,0.58,0.82}
\lstdefinelanguage{Dafny}{
  keywords={method,function,predicate,lemma,class,datatype,trait,var,ghost,returns,modifies,reads,if,then,else,while,forall,exists,in,returns,true,false,null,this,new
  },
  sensitive=true,
  comment=[l]{//},
  morecomment=[s]{/*}{*/},
  morestring=[b]",
}
\newcommand{\Nat}{\mathbb{N}}
\newcommand{\commentout}[1]{}

\newcommand{\tc}[1]{\texttt{\small #1}}

\newcommand{\set}[1]{\{#1\}}

\newcommand{\equivX}[1]{\stackrel{#1}{\sim}}
\renewcommand{\S}{\mathbb{S}}

\title{Verifying Type Soundness for Nontransitive Information Flow Security} 
\author{Chenyi Zhang}
\institute{Computer Science and Software Engineering, University of Canterbury\\
Christchurch, New Zealand}

\begin{document}

\maketitle

\begin{abstract}
    We present a machine-checked soundness proof of a type system for the nontransitive noninterference (NTNI) information flow security property in an imperative programming language using Dafny. The original formulation of NTNI was developed for a functional programming language augmented with memory access operators, where security policies are expressed as coarse-grained information flow constraints over component-based software systems. Although subsequent research has extended NTNI type systems in various directions, to the best of our knowledge, no prior work has established a machine-checked soundness proof for type-enforced NTNI. As a result, the correctness of existing NTNI enforcement mechanisms has not been formally verified with a high degree of assurance.

    Dafny is a verification-oriented programming language and proof assistant that supports reasoning about functional, imperative, and object-oriented programs. In this work, we formalize the operational semantics, security policies, and typing rules of the language within Dafny’s term algebra framework. Based on these formalizations, we mechanically verify that every program accepted by the type system satisfies the intended NTNI security guarantees.
\end{abstract}

\section{Introduction}\label{sec:introduction}

Language-based security leverages program semantics and static program analysis to enforce security properties. One of the most extensively studied classes of such properties is information flow security, which aims to ensure the confidentiality and integrity of programs~\cite{sabelfeld2003language}. Since the 1990s, numerous formal approaches based on secure type systems and Hoare logic-style proof systems have been proposed for imperative~\cite{volpano1996sound}, functional~\cite{heintze1998slam}, and object-oriented~\cite{myers1999jflow} programming languages, supporting both strict noninterference and controlled information downgrading policies~\cite{sabelfeld2009declassification}.

Most language-based information flow control (IFC) systems build upon Denning’s lattice model~\cite{denning76,denning77}, which assumes transitive information flow policies. In these systems, security levels form a partially ordered lattice that is used to classify program entities such as inputs, outputs, and intermediate variables, thereby enabling the tracking of information dependencies throughout program execution. However, in some settings, the transitivity assumption is overly restrictive or undesirable.
Modern software systems are increasingly composed of reusable components obtained from diverse and potentially untrusted sources. Such components, including third-party code downloaded from the Internet, may execute alongside sensitive system components within the same process. In these environments, security policies cannot rely solely on the intentions of the original software developers. Instead, application developers, system integrators, or deployers must be able to specify their own security 
requirements to protect trusted system components from untrusted code.
Since
users often lack knowledge of the internal implementation details of third-party components, they can typically specify only coarse-grained security policies based on the intended functionality of each component. Consequently, security labels are associated with 
software components rather than with individual program variables. To address these requirements, nontransitive noninterference (NTNI)~\cite{lu2020nontransitive} was proposed as a more flexible foundation for secure software composition.

\begin{figure}[!t]
\begin{lstlisting}[%caption=code 1,frame=tlrb
language = Java, keywordstyle=\bf, basicstyle=\ttfamily\small, commentstyle=\ttfamily\color{gray}, xleftmargin=1.2em, numbers=none, firstnumber = 1, escapeinside={(*@}{@*)}]{}
Library { // protected code: only accessible via Service
  process(source_data){ ... }
  retrieve(key){
    ...
    return some_data;
  }
}
Service { // privileged code
  logFile;  
  addLog(x) { 
    append(logFile, x);
  }
  print(data) { 
    Library.process(data);
  }
  query_data(key){
    return Library.retrieve(key);
  }
}
Downloaded_Code { // untrusted code
  v, d = ...;     // untrusted data
  main() {
    Service.addLog(d); 
    Service.print(d.data);
    v = Service.query_data(d.key);  
  } 
}
\end{lstlisting}
\caption{Confused deputy attack.\label{fig:cda}}
\end{figure}

In the example shown in Fig.~\ref{fig:cda}, we assume a local system in which the \tc{Library} is protected and accessible via only the privileged code \tc{Service}.
Code downloaded from Internet is not trusted, which may be permitted to access \tc{Service} but not \tc{Library}. 
For instance, \tc{Downloaded\_Code} is allowed to invoke \tc{addLog()} that appends some (untrusted) data into a non-executable log file to be inspected manually by the system administrator. 
Since the appended data is not further propagated to other parts of the system, this action is regarded as \emph{secure}. 
The untrusted \tc{Downloaded\_Code} also invokes \tc{print()} from \tc{Service} that reads and prints out data using protected \tc{Library} resources via \tc{process()}. 
This call is \emph{insecure}.
Although the function \tc{print()} has the privilege to call \tc{process()} from \tc{Library}, and \tc{Downloaded\_code} has the privilege to use the \tc{Service} interface, \tc{Downloaded\_Code} should not influence the execution of \tc{Library} functions. 
Consequently, this exhibits a \emph{confused deputy problem}---while \tc{Downloaded\_code} cannot directly access the printout in \tc{Library}, it can still influence the printout via the (confused) deputy \tc{Service}.\footnote{This issue can also be identified via the dynamic \emph{stack inspection} mechanism~\cite{popl02,wallach1998understanding}, though there are subtle differences between the approaches of stack inspection and information flow, with various examples discussed in this report~\cite{lu2016unifying}.}

Transitive noninterference with fine-grained security policies can solve the above problem by assigning security levels to individual program variables in \tc{Service}, for example, specifying \tc{addLog.x} to be accessible to all code while \tc{print.data} to be accessible to only \emph{trusted} local code.
However, since \tc{Service} may contain a large number of variables and functions to be shared among a variety of third party applications, specification of individual program variables may be difficult and error-prone. 
NTNI with coarse-grained security policies provides a simpler and more intuitive solution, by assigning security levels $L$, $S$ and $D$ to code bases \tc{Library}, \tc{Service} and \tc{Downloaded\_Code} respectively, and specifying $D\unrhd S$, $S\unrhd L$,  $L\unrhd S$ and $S\unrhd D$ (disallowing $D\unrhd L$ and $L\unrhd D$).
With such a policy, \tc{Service.query\_data()} will be a violation of the policy as it retrieves information from \tc{Library} via \tc{Service}, and will be rejected given $L\unrhd D$ disallowed.
%
%

Type systems have been widely adopted for the enforcement of information flow security properties in programming languages, such as in~\cite{sabelfeld2003language,sabelfeld2009declassification}. However, most existing approaches do not provide machine-checkable proofs of soundness, with a few notable exceptions using Coq~\cite{kammuller2008formalizing} and Isabelle/HOL~\cite{murray2013sel4}. In particular, Murray et al.~\cite{murray2013sel4} conducted a machine-checked proof that the seL4 microkernel enforces information flow security, proving a variant of intransitive noninterference for the actual C implementation of seL4. In another project, Kamm\"uller~\cite{kammuller2008formalizing} formalizes a simple bytecode language and proves noninterference by leveraging Coq’s module system. 
The type system for NTNI 
has only a pen-and-paper proof for the soundness available in~\cite{lu2020nontransitive}. 


This work presents a machine-checked soundness proof of a type system for nontransitive noninterference (NTNI) implemented in Dafny~\cite{leino2010dafny}. In contrast to the substantial proof engineering effort typically required in interactive theorem provers such as Coq and Isabelle/HOL, Dafny provides a lightweight framework for the formalization and verification of programming language semantics and type systems. Specifications, predicates, and rewriting rules encoded in Dafny are automatically translated into verification conditions discharged by backend SMT solvers, significantly reducing the manual effort required for mechanized proofs. As a result, Dafny enables type systems to be formalized and verified in a relatively concise, structured, and accessible manner.

Furthermore, unlike the original NTNI formulation in~\cite{lu2020nontransitive}, which was presented in a functional programming setting, we express NTNI over a while-language similar to that used in the Volpano–Smith–Irvine (VSI) type system~\cite{volpano1996sound}, the first sound type system for transitive noninterference. By adopting this simpler imperative setting, we aim to make the intuition behind the correctness of the NTNI type system clearer and more accessible than in the original presentation.

Our Dafny formalization is publicly available\footnote{The code is located at \url{https://github.com/czhang-fm/TypeSoundnessNTNI/}}.
The language semantics, type system and soundness proof are implemented in compact Dafny modules, which consist of $635$ lines of Dafny code, including comments, sanity tests and assertions. (The total is $195$ for the language model, $180$ for the type system and $260$ for the soundness proof.)
This is a manageable size for a simple language with mostly quantifier-light arithmetic and set-theoretic reasoning, which makes it ideal for the SMT backend to verify the proof without imposing too much additional trust.
For comparison the Coq model for (transitive) non-interference in a simple bytecode language requires approximately $2,300$ lines of code~\cite{kammuller2008formalizing}.

\smallskip

Our contributions are summarized as follows.
\begin{enumerate}
    \item We present a machine-checked soundness proof for a type system enforcing nontransitive noninterference (NTNI) based on the while-language of~\cite{volpano1996sound,smith2007principles}. 
    As a byproduct, we also provide a Dafny mechanization of the soundness proof for the Volpano–Smith–Irvine (VSI) type system, whose original presentation relied solely on pen-and-paper proofs. Note we are \emph{not} the first to provide a machine-checked proof for the VSI type system, as there exist known works such as~\cite{VolpanoSmith-AFP} with a proof in Isabelle. 
    \item Our proof of the NTNI type system clarifies that, for each security level $\ell$, the corresponding security reasoning can be reduced to a two-level information flow security problem over an extended set of program variables. This reduction is established directly from the typing derivation of a program and therefore applies only to well-typed programs. Compared with the approach of Ahmadpanah et al.~\cite{ahmadpanah2021nontransitive}, which reduces NTNI to transitive noninterference through program transformation and potentially increases the size and complexity of the program, our reduction is conceptually simpler and more~direct. 
\end{enumerate}


\section{A System Model}\label{sec:system}
We encode the security condition in the imperative programming language specified in Fig.~\ref{fig:language}, where metavariable $x$ ranges over variables and $v$ over integer values. We use $0$ for false and nonzero values for true. A command is also known as a program. Given a program, $V$ is the set of variables that appear in the~program.

\begin{figure}[!t]
\centering
\begin{tabular}[c]{llr}
		$e::=$ &  & \emph{Expressions} \\
		& $v$ & value\\
		& $x$ & variable\\
		& $e + e$ & binary operation\\
        $c::=$ & & \emph{Commands}\\
        & $\mathtt{skip}$ & termination\\
		& $x:=e$ & assignment\\
		& $\mathtt{if} \ e \ \mathtt{then} \ c_1 \ \mathtt{else} \ c_2$ & conditional\\
		& $\mathtt{while} \ e \ \mathtt{do} \ c $ & loop\\
        & $c_1;c_2$ & sequential composition
	\end{tabular}
	\caption{Language syntax. \label{fig:language}}
\end{figure}

A program $c$ is executed under a memory state $M:V\to \Nat$ which assigns a value to each variable, with a small step semantics defined in Fig.~\ref{fig:semantics}, following the standard structural operational semantics as in Gunter~\cite{gunter1992semantics}, where $(M, c)\to (M', c')$ indicates that program $c$ under memory state $M$ performs a single step execution and then becomes $c'$ with the memory state updated to $M'$.

\begin{figure*}[!t]
	\[
        \frac{M \ e_1 \Downarrow v_1\qquad M \ e_2 \Downarrow v_2}{M\ e_1 + e_2 \Downarrow v_1+v_2}
        \quad \text{[EXP]}
        \qquad
        \frac{}{M\ v \Downarrow v} 
        \quad \text{[VAL]} 
	\]
    \[
        \frac{
			M(x) = v
		}{
			M \ x \Downarrow v
		}
		\quad \text{[VAR]} \qquad
        \frac{
			M \ e  \Downarrow v
		}{
			(M, x {:=} e) \to (M[x\mapsto v], \mathtt{skip}) 
		}
		\quad \text{[ASN]}
    \]
	\[
		\frac{
			(M, c_1) \to (M', c_1')
		}{ 
			(M, c_1; c_2) \to (M', c_1';c_2) 
		}
		\quad \text{[SEQ1]}
        \qquad
        \frac{
            (M, c_2) \to (M', c_2')
        }{
            (M, \mathtt{skip}; c_2) \to (M', c_2')
        }
        \quad \text{[SEQ2]}
	\]
    \[
        \frac{
            M \ e \Downarrow v \quad v\neq 0
        }{
            (M, \mathtt{while} \ e \ \mathtt{do} \ c) \to (M, c ; \mathtt{while} \ e \ \mathtt{do} \ c)
        }
        \text{[WHT]}
        \quad
        \frac{
            M \ e \Downarrow 0
        }{
            (M, \mathtt{while} \ e \ \mathtt{do} \ c) \to (M, \mathtt{skip})
        }
        \text{[WHF]}
    \]
	\[
		\frac{
			M \ e   \Downarrow v \quad v\neq 0 
		}{ 
			(M, \mathtt{if} \ e \ \mathtt{then} \ c_1 \ \mathtt{else} \ c_2)  \to (M, c_1)
		}
		\text{[TRU]} \quad 
		\frac{
			M \ e   \Downarrow 0 
		}{ 
			(M, \mathtt{if} \ e \ \mathtt{then} \ c_1 \ \mathtt{else} \ c_2)  \to (M, c_2)
		}	
		\text{[FLS]}
	\]
	\caption{Structural operational semantics.}
	\label{fig:semantics}
\end{figure*}

The VSI type system~\cite{volpano1996sound} enforces the two level security policy $L\unrhd H$ assuming there exists a partition over program variables into $L$ and $H$, and the information stored in variables labeled with security level $L$ is permitted to flow to variables with security $H$, but not the other way around. A more precise way of defining this security property is by adopting the notion of \emph{noninterference} which is initially proposed by Goguen and Meseguer~\cite{goguen_security_1982}. In the programming language setting, we need a notion of equivalence over memory states to express this property. Define $M_1=_L M_2$ if two memory states $M_1$ and $M_2$ agree on the values for all variables labeled by $L$ (so called \emph{low-equivalence states}), and ``$\rightarrow^*$" is the transitive closure of the small-step transition relation ``$\rightarrow$" defined in Fig.~\ref{fig:semantics} provided that an execution terminates. 
\begin{definition}[Termination-Insensitive Noninterference, TI-NI]\label{def:ni}
Program $c$ satisfies TI-NI wrt to the policy $L\unrhd H$, if for all memory states $M_1$ and $M_2$, we have 
\[
	\left.
		\begin{array}{c}
			M_1 =_L M_2
			\\	
			(M_1, c)\rightarrow^*\ (M_1',\mathtt{skip})
			\\
			(M_2, c)\rightarrow^*\ (M_2',\mathtt{skip})
		\end{array}
	\right\}
		\implies
		M_1' =_L M_2'
\]
\end{definition}
We note that Definition~\ref{def:ni} is termination insensitive, as similar to~\cite{volpano1996sound}, this paper does not consider information leakage caused by program termination. 
Consequently, we often omit ``TI" from definitions if it is clear from the context. 


\paragraph{Explicit flow.} Explicit information flow occurs when a variable with a high security level appears directly in an expression assigned to a variable with a lower security level, such as in the assignment $l := h + 2$, where $l$ is labeled $L$ and $h$ is labeled $H$. 
Explicit flow patterns also arise in input/output operations and in value-returning function or method calls. 

\paragraph{Implicit flow.} Implicit information flow occurs when a high-security variable influences the control flow of a program, thereby affecting the value of a low-security variable indirectly. A typical example is the statement $\mathtt{if} \ h \ \mathtt{then} \ l := 1 \ \mathtt{else} \ l:=0$, where $h \in \{0,1\}$. In this case, the value of $h$ is effectively copied to $l$ through the branching structure. 

\medskip

\begin{figure*}[t]
	\[
        \frac{\Sigma \vdash e_1:\tau\qquad \Sigma \vdash e_2:\tau}{\Sigma\vdash e_1 + e_2 :\tau}
        \quad \text{[TEXP]}
        \qquad
        \frac{}{\Sigma \vdash v:L} 
        \quad \text{[TVAL]} 
	\]
    \[
        \frac{
			\Sigma(x) = \tau\ \texttt{var}
		}{
			\Sigma \vdash x : \tau
		}
		\quad \text{[TVAR]} \qquad
        \frac{
			\Sigma \vdash x:\tau \quad \Sigma\vdash e :\tau
		}{
			\Sigma\vdash x {:=} e :\tau 
		}
		\quad \text{[TASN]}
    \]
	\[
		\frac{
			\Sigma\vdash c_1\ \tau\ \mathtt{cmd} \quad \Sigma\vdash  c_2\ \tau\ \mathtt{cmd}
		}{ 
			\Sigma\vdash c_1; c_2: \tau  \ \mathtt{cmd}
		}
		\quad \text{[TSEQ]}
        \qquad
        \frac{
            \Sigma\vdash e:\tau\quad \Sigma\vdash c:\tau\ \mathtt{cmd}
        }{
            \Sigma\vdash \mathtt{while} \ e \ \mathtt{do} \ c: \tau\ \mathtt{cmd}
        }
        \quad \text{[TWH]}
    \]
	\[
		\frac{
			\Sigma\vdash e:\tau\quad \Sigma\vdash c_1:\tau\ \mathtt{cmd}\quad \Sigma\vdash c_2:\tau\ \mathtt{cmd}
		}{ 
			\Sigma\vdash \mathtt{if} \ x \ \mathtt{then} \ c_1 \ \mathtt{else} \ c_2:\tau\ \mathtt{cmd}
		}
		\quad \text{[TIF]}
	\]
    \[
    \frac{}{L\unrhd H}\quad\text{[BASE]}\qquad\frac{\tau_1\unrhd\tau_2}{\tau_2\ \mathtt{cmd}\unrhd\tau_1\ \mathtt{cmd}}\quad\text{[CMD]}\qquad\frac{}{\rho\unrhd \rho}\quad\text{[REFL]}
    \]
    \[
    \frac{\rho_1\unrhd\rho_2\quad \rho_2\unrhd\rho_3}{\rho_1\ \unrhd\rho_3}\quad\text{[TRANS]}\qquad\frac{\Sigma\vdash c:\rho_1\quad \rho_1\unrhd\rho_2}{\Sigma\vdash c:\rho_2}\quad\text{[SUB]}
    \]
	\caption{The Volpano-Smith-Irvine type system.}
	\label{fig:vsi-type}
\end{figure*}

We briefly review the VSI type system defined in Fig.~\ref{fig:vsi-type}. 
The VSI system defines two base security types, $L$ and $H$, under the security policy $L \unrhd H$. The variable type $\tau\ \mathtt{var}$ associated with a variable $x$ indicates that $\tau$ is the security type assigned to $x$, as specified by the programmer or user. The type $\tau\ \mathtt{cmd}$ is used exclusively for typing commands (or programs).
The key insight of the VSI type system is that an expression is always assigned the highest security level among the information it depends on, whereas a command $c$ is assigned type $\tau\ \mathit{cmd}$, where $\tau$ is the lowest security level among the variables that may be updated during the execution of $c$. 


\begin{theorem}[VSI soundness~\cite{volpano1996sound}]\label{thm:ni}
If $c$ is well typed in VSI, then $c$ satisfies noninterference with respect to the policy $L\unrhd H$.
\end{theorem}

It is sometimes argued that given a multi-level security policy $(\S, \unrhd)$, where $\S$ is a set of security levels and $\unrhd$ a partial order over $\S$, 
security reasoning can be reduced to a collection of two-level security problems, one for each security level $\ell \in \S$.
In particular, for a fixed level $\ell$, let $L_\ell = \set{\ell'\mid \ell'\unrhd\ell}$ and $H_\ell = \S\setminus L_\ell$. 
Security is then verified by checking that information from $H_\ell$ does not interfere with information observable at levels in $L_\ell$ for every security level $\ell \in \S$.


\section{Nontransitive Noninterference}\label{section:noninterference}
In a system governed by a transitive information flow policy, information associated with a particular security level may flow to any other security level reachable through the transitive closure of the policy relation. This characteristic can make transitive policies unsuitable for expressing coarse-grained security requirements in component-based systems.

For example, suppose a component owned by Alice is willing to share information only with another component, Bob. Under a transitive policy, if Bob is permitted to share information with additional components, then Alice’s information may also flow indirectly to those components through the transitive closure of the policy relation. Such indirect flows may violate Alice’s intended security requirements and are therefore undesirable.


The above requirement leads us to a scheme that defines a flow policy in a less permissive way by removing the transitivity requirement from the policy $\unrhd$ which is a binary relation over the set of security levels.
Taking the policy required for the example in Fig.~\ref{fig:cda}, which contains code bases $\tc{Library}\ (L)$ (protected code), $\tc{Service}\ (S)$ (privileged code), and $\tc{Download\_Code}\ (D)$, where the security levels are in $\S = \set{L, S, D}$ with the permitted flow relation $D\unrhd S$, $S\unrhd L$,  $L\unrhd S$ and $S\unrhd D$ (i.e., $D\unrhd L$ and $L\unrhd D$ are disallowed). Take $\tc{Download\_Code}\ D$, the security policy requires that the final values of variables typed $D$ should only be influenced by input values to variables typed either $D$ or $S$ (Since $S\unrhd D$), but not $L$. This leads us to the following definition.

For each $\ell\in \S$ define $\unrhd(\ell)=\set{\ell'\mid \ell'\unrhd\ell}$, which is the set of security levels that are allowed to influence $\ell$. Moreover, we assume $\unrhd$ is reflexive, thus $\ell\in\unrhd(\ell)$ for any $\ell\in\S$. Write $M_1\equivX{T}M_2$ if for all variables $x$ that are assigned security level $\ell\in T$, we have $M_1(x)=M_2(x)$. 

\begin{definition}[Termination-Insensitive Nontransitive Noninterference, TI-NTNI]\label{def:ntni}\rm{\cite{lu2020nontransitive}}
Program $c$ satisfies TI-NTNI with respect to a nontransitive policy $\unrhd$, if for all memory states $M_1$ and $M_2$ and security level $\ell\in \S$, we have 
\[
	\left.
		\begin{array}{c}
			M_1 \equivX{\unrhd(\ell)} M_2
			\\	
			(M_1, c)\rightarrow^*\ (M_1',\mathtt{skip})
			\\
			(M_2, c)\rightarrow^*\ (M_2',\mathtt{skip})
		\end{array}
	\right\}
		\implies
		M_1' \equivX{\set{\ell}} M_2'
\]
\end{definition}

\begin{figure*}[t]
	\[
        \frac{\Sigma, \Gamma \vdash_{pc} e_1:\tau\qquad \Sigma, \Gamma \vdash_{pc} e_2:\tau}{\Sigma, \Gamma\vdash_{pc} e_1 + e_2 :\tau}
        \quad \text{[NEXP]}
        \qquad
        \frac{}{\Sigma, \Gamma \vdash_{pc} v:\emptyset} 
        \quad \text{[NVAL]} 
	\]
    \[
        \frac{
			\Sigma(x) = \ell
		}{
			\Sigma, \Gamma \vdash_{pc} x : \set{\ell}
		}
		\quad \text{[NVAR]} \qquad
        \frac{
			\Sigma,\Gamma\vdash_{pc} e :\tau\quad \tau\cup pc\subseteq \Gamma(x)\subseteq\unrhd(\Sigma(x))
		}{
			\Sigma,\Gamma\vdash_{pc} x {:=} e :\set{x} 
		}
		\quad \text{[NASN]}
    \]
	\[
		\frac{
			\Sigma,\Gamma\vdash_{pc} c_1: \rho \quad \Sigma,\Gamma\vdash_{pc}  c_2: \rho
		}{ 
			\Sigma,\Gamma\vdash_{pc} c_1; c_2: \rho  
		}
		\quad \text{[NSEQ]}
        \qquad
        \frac{
            \Sigma,\Gamma\vdash_{pc} e:\tau\quad \Sigma,\Gamma\vdash_{pc\cup\tau} c:\rho
        }{
            \Sigma,\Gamma\vdash_{pc} \mathtt{while} \ e \ \mathtt{do} \ c: \rho
        }
        \quad \text{[NWH]}
    \]
	\[
		\frac{
			\Sigma,\Gamma\vdash_{pc} e:\tau\quad \Sigma,\Gamma\vdash_{pc\cup\tau} c_1:\rho \quad \Sigma,\Gamma\vdash_{pc\cup\tau} c_2:\rho
		}{ 
			\Sigma,\Gamma\vdash_{pc} \mathtt{if} \ x \ \mathtt{then} \ c_1 \ \mathtt{else} \ c_2:\rho
		}
		\quad \text{[NIF]}
	\]
    \[
    \frac{\Sigma, \Gamma \vdash_{pc} e:\tau_1\quad \tau_1\subseteq\tau_2}{\Sigma, \Gamma \vdash_{pc} e:\tau_2}\quad\text{[E-SUB]}
    \qquad
    \frac{\Sigma, \Gamma \vdash_{pc} c:\rho_1\quad \rho_1\subseteq\rho_2}{\Sigma, \Gamma \vdash_{pc} c:\rho_2}\quad\text{[C-SUB]}
    \]
	\caption{The nontransitive noninterference type system.}
	\label{fig:ntni-type}
\end{figure*}

The type system described in Fig.~\ref{fig:ntni-type} enforces a nontransitive noninterference policy $\unrhd$ over a set of security levels $\S$.
Type context $\Sigma$ assigns a security level $\Sigma(x)\in \S$ to each variable $x$ in the program, and type context $\Gamma$ 
assigns each variable $x$ a collection of security levels that may (transitively) flow to $x$. 
In particular, for each expression $e$, the type system assigns it a type $\tau$ that contains all information that may influence the value of $e$, and for each command $c$, a command type $\rho$ is the collection of variables that may get updated in the execution of $c$. In particular, \text{[C-SUB]} says if $\rho_1$ is the set of variables that may get updated ($\Sigma,\Gamma\vdash_{pc}c:\rho_1$), then it is safe to say that $c$ can be typed with a larger set (i.e., $\Sigma,\Gamma\vdash_{pc}c:\rho_2$ with $\rho_1\subseteq\rho_2$). Likewise, for \text{[E-SUB]}, if an expression $e$ can be typed with a set of labels $\tau_1$, then it can also be typed $\tau_2$ given $\tau_1\subseteq\tau_2$. Now the subtyping relation is just the subset relation, so the reflexivity and transitivity typing rules that appears in the VSI type system are no longer required in the type system for NTNI.\footnote{Arbitrarily enlarging the type for a particular expression or statement will be constrained in the process of establishing a type for the entire program.} In those rules, control dependencies are carried by $pc$, which is propagated via rules such as \text{[NIF]} and \text{[NWH]}, and enforced by \text{[NASN]} when an assignment is located inside a branching structure or a loop structure. In the Dafny model, the command type $\rho$ is used to indicate whether the provided pair of type contexts $\Sigma$ and $\Gamma$ are valid for a program, so that invalid type contexts should set $\rho$ to \texttt{invalid}.


Perhaps the most important constraint is defined in the rule \text{[NASN]} which assigns a type to statement $x {:=} e$. In this case, we need to ensure that all information contained in the right-hand-side (i.e., $e:\tau$) as well as information for control dependencies (i.e., $pc$) be allowed to influence $x$ (i.e., $pc\subseteq\unrhd(\Sigma(x))$. The same rule also enforces $\Gamma(x)\subseteq\unrhd(\Sigma(x))$ at the global level. The information encoded in $pc$ about control dependencies is propagated via the rules \text{[NIF]} and \text{[NWH]} from upper level syntactic structures.

The type system introduced in Fig.~\ref{fig:ntni-type} enforces NTNI. This theorem is proved in the Dafny model that is to be introduced in the following section.~\footnote{For a program $c$, we only require the existence of type contexts $\Sigma$ and $\Gamma$ to establish NTNI for $c$. Such a type system is usually incomplete, i.e., there may exist a program that satisfies NTNI but cannot be typed in the system defined in Fig.~\ref{fig:ntni-type}. Computing type contexts from a program and a policy is also out of the scope of this paper.}
\begin{theorem}[NTNI soundness]\label{thm:ntni}
If $c$ is well typed in the NTNI type system with respect to a security policy $\unrhd$, then $c$ satisfies nontransitive noninterference with respect to $\unrhd$.
\end{theorem}
\section{Dafny Encoding of NTNI}\label{sec:dafny}
Dafny is a proof assistant that supports both functional programming styled specifications and imperative programming styled specifications~\cite{leino2010dafny}. A lemma or theorem in Dafny is essentially a method that enforces whenever 
its pre-conditions are satisfied, 
all its post-conditions be satisfied. This basic while-programming language is modeled with a user-defined (variant) type as follows.
\smallskip

\begin{lstlisting}[caption={Program syntax}, label={lst:syntax}]
    // The type for all expressions 
    datatype Expr = | Num(n: int) | Var(x: Variable) 
                    | Plus(e1: Expr, e2: Expr)
    // The type for all commands (statements)
    datatype Cmd = | Skip | Assn(x: Variable, e: Expr) 
                   | If(cond: Expr, thn: Cmd, els: Cmd) 
                   | While(cond: Expr, c: Cmd) 
                   | Seq(c1: Cmd, c2: Cmd)
\end{lstlisting}  


\begin{lstlisting}[caption={Structure Operational Semantics}, label={lst:sos}]
function SmallStep(s: MState, c: Cmd) : (MState, Cmd) 
// ...
{
    match c {
        case Skip =>  (s, c) 
        case Assn(x, e) => (s[x:= Evaluate(s, e)], Skip) 
        case If(e, c1, c2) => // if-then-else
            var res := Evaluate(s, e);
            if (res != 0) then (s, c1) else (s, c2)
        case While(e, c1) => // while-loop
            var res := Evaluate(s, e);
            if (res != 0) then (s, Seq(c1, c)) 
                else (s, Skip)
        case Seq(c1, c2) => // sequential composition
            match c1 {
                case Skip => (s, c2) // removing 1st skip
                case c1 => 
                    var (s', c') := SmallStep(s, c1); 
                    (s', Seq(c', c2)) }  
    }
}
\end{lstlisting}

The structural operational semantics (SOS) define how such a program is executed, which is essentially modeled as a function that reads memory state $s$ and command $c$ as inputs, then performs a single step execution that may update memory state $s$ into a new state with a (likely) shorter command. Inside the execution model, $\texttt{Evaluate(s, e)}$ evaluates expression $\texttt{e}$ according to memory state $\texttt{s}$; If $\texttt{c}$ is a sequential composition, then the function may recursively invoke itself as per the SOS rule $\text{[SEQ1]}$ in Fig.~\ref{fig:semantics}.

The type checking procedure is modeled with two functions. The first function, $\texttt{HasExprType(vctx, e)}$, gathers information (the set of security labels) that flows into expression $\texttt{e}$. In particular, the amount of information represented by variable $x$ is the $x$ itself together with the information flowed into $x$ ($\texttt{vctx[x]}$). The information that flows into $\texttt{e1} + \texttt{e2}$ is recursively defined as the join of the information that flows into $\texttt{e1}$ and the information that flows into $\texttt{e2}$, where the \texttt{Join} operator is essentially set union.  
\smallskip

\begin{lstlisting}[caption={Expression Type Checking (simplified for readability)}, label={lst:expr_type}]
function HasExprType(vctx:VContext, e:Expr): PhraseType
{
    match e {
        case Num(n) => ExprType({})
        case Var(x) => ExprType(vctx[x]+{x}) 
        case Plus(e1, e2) => 
            var t1 := HasExprType(vctx, e1); 
            var t2 := HasExprType(vctx, e2); 
            ExprType(Join(t1, t2))
    }
}
\end{lstlisting}

The $\texttt{HasCmdType(vctx, vpc, c)}$ function checks if $c$ is typable under the context $\texttt{vctx}$ with the control dependencies represented by \texttt{vpc}. If type checking succeeds, the set of potentially updated variables within $\texttt{c}$ will be returned; otherwise, it returns $\texttt{Invalid}$. Any command with an $\texttt{Invalid}$ component is also not typable.
We focus only on the assignment and the if-condition cases, as these two structures should be sufficient for explaining how $\texttt{HasCmdType()}$ works. For $\texttt{Assn(x, e)}$, the security level for $x$ is stored at $\texttt{t}_1$ and the information from $\texttt{e}$ stored at $\texttt{t}_2$ with labels in $\texttt{e\_labels}$. The model then enforces that the security labels stored at $\texttt{t}_2$ (i.e., $\texttt{e\_labels}$) should be a subset of $\texttt{CanFlow(t1)}$ which is an upper bound for information allowed to enter $x$ as per the security policy. Variables 
for control dependencies ($\texttt{pc\_labels}$) are also checked against $\texttt{CanFlow(t1)}$. In the case of $\texttt{If(e, c1, c2)}$, both branches $\texttt{c1}$ and $\texttt{c2}$ are required to be checked with a (possibly larger) new control dependencies set ($\texttt{vpc+te}$) by including the new information from $\texttt{e}$ via~$\texttt{te}$.

\smallskip

\begin{lstlisting}[caption={Command Type Checking (simplified for readability)}, label={lst:cmd_type}]
function HasCmdType(vctx:VContext,vpc:set<Variable>,c:Cmd): PhraseType 
// ...
{
    match c {
      case Skip => CmdType({})
      case Assn(x, e) =>  
        // t1 is the policy label of x
        // t2 represents information from the RHS of =
        var t1, t2 := label[x], HasExprType(vctx, e); 
        // get the set of labels allowed to influence x
        var flowsFrom := CanFlow(t1);
        var pc_labels := (set x | x in vpc :: label[x]);
        var e_labels := (set x | x in t2 :: label[x]);
        if (t2+vpc+{x}) <= vctx[x] 
            && (e_labels+pc_labels) <= flowsFrom
        then CmdType({x}) 
        else Invalid
      case If(e, c1, c2) => 
        var te := HasExprType(vctx, e);
        var t1 := HasCmdType(vctx, vpc+te,c1); 
        var t2 := HasCmdType(vctx, vpc+te,c2);
        if t1 == Invalid || t2 == Invalid then Invalid
        else (CmdType(Join(t1, t2)))
      case While(e, c1) => //...
      case Seq(c1, c2) => ///...
    }
}
\end{lstlisting}

\begin{lstlisting}[caption={NTNI security by the type system (Theorem $2$), simpliefied for readability}, label={lst:NTNI}]
lemma Noninterference(vctx: VContext, ... s1, s2, s1', s2': State) // other preconditions...
requires HasCmdType(vctx, {}, c) != Invalid 
// type checking passed
requires Equiv(vctx, (set x | x in variables && coarse_label[x] in CanFlow(l) :: x), s1, s2)
requires Terminates(c,s1,s1') && Terminates(c,s2,s2')
ensures Equiv(vctx, (set x | x in variables && coarse_label[x] == l::x),s1',s2')
\end{lstlisting}

Listing~\ref{lst:NTNI} is a simplified Dafny representation for Theorem~\ref{thm:ntni}. Here, \texttt{(set x | x in variables \&\& coarse\_label[x] in CanFlow(l) :: x)} represents the set of variables with labels in $\unrhd(\ell)$ and the notation \texttt{(set x | x in variables \&\& coarse\_label[x] == l::x)} is the set of variables $x$ with $\Sigma(x)=\ell$ (i.e., \texttt{coarse\_label[x]} retrieves the assigned security label for $x$). In the following, we discuss how to establish a proof for Theorem~\ref{thm:ntni}.

\subsubsection{Towards a soundness proof for NTNI}

The soundness of VSI type system for the two-level security policy $L\unrhd H$ (Theorem~\ref{thm:ni}) is established by showing that the equivalence relation $=_L$ can be maintained on a pair of initial states $s_1$ and $s_2$ when they both go along with all the execution steps in a well-typed program $c$. This then requires us to check all possible constructions of program $c$ via structural induction. 

This proof technique, however, cannot be directly applied to the case of nontransitive noninterference (Theorem~\ref{thm:ntni}). The assumption of Definition~\ref{def:ntni} requires two memory state agree on all values stored in security levels $\unrhd(\ell)$, 
which is often a larger set of variables than those have security level $\ell$ which appears in the conclusion of Definition~\ref{def:ntni}. 
Given $\equivX{\unrhd(\ell)}\subseteq\equivX{\set{\ell}}$, 
a possible solution is to find a new equivalence relation on states which sits in between $\equivX{\unrhd(\ell)}$ and $\equivX{\set{\ell}}$ and works in the same way as the low equivalence relation ($=_L$) in the proof of Theorem~\ref{thm:ni}. Provided that type checking succeeds, such a relation can be constructed from the following set of variables: $\set{ y\in V\mid \exists x.\Sigma(x)=\ell\wedge y\in\hat{\Gamma}(x)}$, where $\hat\Gamma(x)$ 
is a (specific) function that contains
all variables that may (transitively) influence each variable $x$ during the type checking, represented by the typing context $\texttt{vctx}$ in Listing~\ref{lst:cmd_type}.~\footnote{Different from $\Sigma$ which can be uniquely determined by a program and a security policy, there may be more than one candidate for $\hat\Gamma$, which makes it possible to have multiple valid pairs of type contexts to satisfy the constraints listed in Fig.~\ref{fig:ntni-type}.}

For each $\ell\in S$, define $Max(\ell)=\set{x\in V\mid \Sigma(x)=\ell}\cup\set{y\in V\mid\exists x.\Sigma(x)=\ell\wedge y\in\hat\Gamma(x)}$. Given memory states $M_1$, $M_2$ and $V'\subseteq V$, write $M_1\equivX{V'}M_2$ if $M_1(x)=M_2(x)$ for all variables $x\in V'$. We establish the following properties that are essential to our proof.

\begin{enumerate}
    \item For all $x$, $y$, $z$, $x\in\hat\Gamma(y)$ and $y\in\hat\Gamma(z)$ implies $x\in\hat\Gamma(z)$;
    \item $\set{x\in V\mid \Sigma(x)=\ell}\subseteq Max(\ell) \subseteq\set{x\in V\mid \Sigma(x)\unrhd\ell}$.
\end{enumerate}
The first property holds by the fact that type checking collects a binary flow relation on variables that is always transitive, established via the construction of $\texttt{ExprType}$ (line~$8$ of Listing~\ref{lst:expr_type}) for both $\texttt{t2}$ (line~$9$ of Listing~\ref{lst:cmd_type}) and $\texttt{te}$ (line~$19$ of Listing~\ref{lst:cmd_type}) and enforced at line~$14$ of Listing~\ref{lst:cmd_type}. Therefore, intuitively, all other variables not included in $Max(\ell)$ cannot influence variables in $Max(\ell)$, as presented in the following lemma. A simplified Dafny representation is presented in Listing~\ref{lst:MaxSet} with the above properties for $Max(\ell)$ encoded as preconditions.
\begin{lemma}[transitive noninterference made from type checking]~\label{lem:max}
If $c$ is well typed in the NTNI type system with respect to a security policy $\unrhd$, then for all $\ell\in S$,
\[
	\left.
		\begin{array}{c}
			M_1 \equivX{Max(\ell)} M_2
			\\	
			(M_1, c)\rightarrow^*\ (M_1',\mathtt{skip})
			\\
			(M_2, c)\rightarrow^*\ (M_2',\mathtt{skip})
		\end{array}
	\right\}
		\implies
		M_1' \equivX{Max(\ell)} M_2'
\]
\end{lemma}
Lemma~\ref{lem:max} and the second property above for $Max(\ell)$ enable a straightforward derivation of Theorem~\ref{thm:ntni}. 

\paragraph{Remark.} The program transformation proposed by Ahmadpanah et al.~\cite{ahmadpanah2021nontransitive} reduces an NTNI security policy to a transitive noninterference policy on a canonical form of the original program by introducing multiple copies of program variables. The construction above suggests that this duplication may not be necessary: for a well-typed program, the type system already induces a partition for each security level $\ell\in S$ that simulates the two-level security policy $(V\setminus Max(\ell))\not\unrhd\ Max(\ell)$.

\smallskip

\begin{lstlisting}[caption={Lemma $1$ simpliefied for readability}, label={lst:MaxSet}]
lemma {:induction false} MaxSetConsistency(vctx: VContext, vset: set<Variable>, ..., s1,s2,s1',s2':State) 
// other preconditions ... 
requires HasCmdType(vctx, vpc, c) != Invalid
requires forall x, y :: x in vset && y in vctx[x] ==> 
    vctx[y] <= vctx[x] <= vset
requires forall x, y :: x in vset && y in vctx[x] ==> 
    coarse_label[y] in CanFlow(coarse_label[x])
requires Equiv(vctx, vset, s1, s2)
ensures Equiv(vctx, vset, s1', s2')
\end{lstlisting}


\section{Related Work}\label{sec:related-work}


The most related works are Lu and Zhang~\cite{lu2020nontransitive}, which introduces the concept of NTNI, and Ahmadpanah et al.~\cite{ahmadpanah2021nontransitive}, which extends NTNI by incorporating flow sensitivity and input-output constructions.  The key concepts discussed in these papers have already been covered in previous sections.


The notion of intransitive noninterference (INI) does not mean that the underlying flow policy is non-transitive; rather, it captures controlled mediation of information flow. Given a policy $A\unrhd B\unrhd C$, information from $A$ may flow to $C$ provided that $B$ acts as a trusted mediator, such as a downgrader or sanitizer. Rushby formalized intransitive noninterference in a state-based setting~\cite{rushby1992noninterference}, introducing unwinding conditions as a proof technique for verification. This notion has been further developed in a range of settings, including trace-based system models~\cite{mantel_controlled_2004} and the process algebra CSP~\cite{roscoe_what_1999}. A related variant of INI was proposed by van der Meyden~\cite{Meyden07,EngelhardtMZ12}, which requires that the downgrader (e.g., $B$) possess at least as much knowledge as the receiver (e.g., $C$) about the information being transferred from the sender (e.g., $A$).
%
These models typically represent policies as graphs over security domains; flows are allowed along certain paths but not others. The formulation of NTNI used in this paper is informed by previous work on coarse-grained nontransitive IFC and its type-theoretic enforcement~\cite{lu2020nontransitive,ahmadpanah2021nontransitive}. 
It may be possible to simulate a policy in INI within the NTNI framework in a way that the designer specifies which edges exist and which paths are legal via a sophisticated policy structure.


Proof assistants like Coq and Isabelle/HOL have been used to formalise proofs of the soundness of type systems and information flow security properties.  For example, Murray et al.~\cite{murray2013sel4} present a machine-checked proof demonstrating that the seL4 microkernel enforces information flow security.  This proof establishes a variant of intransitive noninterference for the actual C implementation of seL4, which comprises approximately $8,830$ lines of code. Another line of research involves Kamm\"uller~\cite{kammuller2008formalizing}, who formalises a simple bytecode language and proves noninterference using Coq’s module system.


Work on downgrading and information declassification forms an orthogonal line of research. Language-based IFC systems such as Jif~\cite{myers1999jflow} and Fabric~\cite{liu_fabric_2017} enforce transitive security policies while accommodating real-world requirements through explicit downgrading primitives, including declassification for confidentiality and endorsement for integrity. DC labels~\cite{StefanRMM11,WBK-stm11} extend this approach with decentralized policies and support for robust declassification and endorsement~\cite{MSZ06,CMA17}. In these systems, downgrading is treated as an explicit operational mechanism that must be carefully introduced and justified by the programmer.




\section{Conclusion}\label{sec:conclusion}


Nontransitive noninterference (NTNI) is a class of security properties useful in settings where information-flow policies are specified in a coarse-grained manner. This work presents a machine-checkable proof of NTNI for a simple imperative programming language, formalized using the Dafny proof assistant. By framing the simplicity of the language as an advantage for pedagogical and mechanization clarity, we show how complex security proofs can be made concise and accessible. The resulting proof development is publicly available. Extending the language and the proof to support full object-oriented programming features, such as classes and method calls, is currently ongoing work. We hope this provides a more solid foundation for existing theoretical developments and for potential future applications~of~NTNI.




\paragraph{Acknowledgment.} The author thanks anonymous reviewers from the program committee of SOFSEM 2027 that helped identify a number of factual and presentation issues in an earlier draft.

\bibliographystyle{plain}
\bibliography{reference}
\end{document}